\documentclass[prl,twocolumn,twoside,superscriptaddress,showpacs,floatfix]{revtex4}
\usepackage{amsmath,graphicx,multirow,natbib,hyperref}
\hypersetup{colorlinks,citecolor=black,filecolor=black,linkcolor=black,urlcolor=black}

\begin{document}

\title{Phase coherence induced by correlated disorder}

\author{Hyunsuk Hong}
\email{hhong@jbnu.ac.kr}
\affiliation{Department of Physics and Research Institute of Physics and Chemistry, Chonbuk National University, Jeonju 561-756, Korea}

\author{Kevin P. O'Keeffe}
\email{kpo24@cornell.edu}
\affiliation{Center for Applied Mathematics, Cornell University, New York 14853, USA} 

\author{Steven H. Strogatz}
\email{shs7@cornell.edu}
\affiliation{Department of Mathematics, Cornell University, New York 14853, USA} 

\date{\today}

\begin{abstract}
We consider a mean-field model of coupled phase oscillators with quenched disorder in the coupling strengths and natural frequencies. When these two kinds of disorder are uncorrelated (and when the positive and negative couplings are equal in number and strength), it is known that phase coherence cannot occur and synchronization is absent. Here we explore the effects of correlating the disorder.  Specifically, we assume that any given oscillator either attracts or repels all the others, and that the sign of the interaction is deterministically correlated with the given oscillator's natural frequency. For symmetrically correlated disorder with zero mean, we find that the system spontaneously synchronizes, once the width of the frequency distribution falls below a critical value. For asymmetrically correlated disorder, the model displays coherent traveling waves: the complex order parameter becomes nonzero and rotates with constant frequency different from the system's mean natural frequency. Thus, in both cases, correlated disorder can trigger phase coherence. 
\end{abstract}

\maketitle


\section{Introduction}

The Kuramoto model has been used to explore the dynamics of synchronization in a wide range of physical, chemical, and biological systems \cite{kuramoto84, strogatz00, pikovsky03, strogatz03, acebron05, dorfler14, rodrigues15}. It has also attracted a great deal of theoretical attention because of the analytical light it sheds on the various collective states and phase transitions that can occur in large systems of coupled nonlinear oscillators.

In Kuramoto's original version of the model \cite{kuramoto84}, the natural frequencies of the oscillators are distributed independently across the population according to a unimodal, symmetric probability distribution with density $g(\omega)$. This disorder in the oscillators' frequencies tends to desynchronize them. Opposing this disorder is an attractive pairwise coupling between the oscillators, of uniform strength $K \ge 0$, which tends to synchronize them. The rich behavior of the model results from the interplay of these competing forces. 

The governing equations of Kuramoto's model are
\begin{equation}
\frac{d\phi_i}{dt} = \omega_i + \frac{K}{N} \sum_{j=1}^{N} \sin(\phi_j-\phi_i), ~~~i= 1,\ldots, N,
\label{eq:kuramoto_model}
\end{equation}
where $\phi_i$ is the phase of the $i$th oscillator and $\omega_i$ is 
its natural frequency. The model's simplifying assumptions of sinusoidal coupling and infinite-range interactions allowed Kuramoto to obtain exact results for its stationary states in the limit $N \rightarrow \infty$, an impressive accomplishment given that the model is, mathematically, an infinite-dimensional nonlinear dynamical system with disorder.  The analysis shows that the model's long-term behavior bifurcates from an incoherent state to a partially coherent state at a critical coupling strength $K_c$. 

In this paper, we consider a variant of the Kuramoto model in which the couplings are distributed as well as the frequencies, but with a correlation between them. The governing equations are now given by
\begin{equation}
\frac{d\phi_i}{dt} = \omega_i + \frac{1}{N} \sum_{j=1}^{N} \xi_j \sin(\phi_j-\phi_i), ~~~i= 1,\ldots, N.
\label{eq:model}
\end{equation}
\noindent
Following Kuramoto \cite{kuramoto84} and many subsequent workers, we assume the frequencies are selected from a Lorentzian distribution given by 
\begin{equation}
g(\omega) = \frac{\gamma}{\pi(\omega^2 + \gamma^2)}, 
\label{eq:lorentzian}
\end{equation} 
having a half-width $\gamma$. The mean frequency is set to zero, $\langle\omega_i\rangle=0$, without loss of generality by going into a suitable rotating frame.
We also introduce disorder in the couplings $\xi_j$ by choosing them from a distribution function $\Gamma(\xi)$.  
For simplicity, we take the double-$\delta$ distribution 
\begin{equation}
\Gamma(\xi) = \frac{1}{2}[\delta(\xi-1)+\delta(\xi+1)],
\label{eq:Gammaxi_delta}
\end{equation}

\noindent
so that half of the oscillators have positive coupling $\xi_j >0$, and the other half have negative coupling $\xi_j< 0$. The positively coupled oscillators attract other oscillators toward them, which promotes global phase coherence. In contrast, negatively coupled oscillators tend to repel others, which inhibits global phase coherence. Since there are equal numbers of positively and negatively coupled oscillators, it is not obvious which of these two competing forces will win out, though incoherence seems likelier, given the additional desynchronizing effects of the random frequencies.

In a previous study~\cite{ref:HS_PRE}, we found that when the random couplings are chosen \emph {independently} of the random frequencies, the system does not show phase coherence.  That result is consistent with what one would expect from a naive mean-field argument:  
If we replace the frequencies and couplings with their zero mean values, $\langle\omega\rangle=0$ and $\langle\xi\rangle=0$, 
there's nothing driving the system toward coherence. 

These observations led us to consider the following questions: Is phase coherence \emph{always} impossible for a symmetric distribution of the disorder in which both the frequencies and couplings have zero mean? What if these two types of disorder are \emph{correlated} rather than independent? Furthermore, what happens if the inhomogeneities in the couplings and frequencies are chosen \emph{deterministically} rather than randomly? The purpose of the present work is to address these questions.

We begin by removing the randomness in the frequencies $\{\omega_i \} $. To do so, we choose the ${\omega_i}$ deterministically such that their cumulative distribution function matches that implied by $g(\omega)$. This condition yields the deterministic frequencies ${\omega_i}$ as the solutions of 
\begin{equation}
\frac{i-0.5}{N}=\int_{-\infty}^{\omega_i} g(\omega) d\omega,
\end{equation}
for $i = 1, \ldots, N$.  For the particular case of the Lorentzian distribution $g(\omega)$ assumed here, this procedure gives
\begin{equation}
\omega_i = \gamma \tan \left[ \frac{i\pi}{N}-\frac{(N+1)\pi}{2N} \right],~~~i=1,\ldots,N. 
\label{eq:regLorw}
\end{equation}

\noindent
Note that since that the frequencies in Eq.~\eqref{eq:regLorw} are deterministic, they comprise only one realization $\{\omega_i\}$.

We next choose the couplings $\{\xi_i \}$. As stated, we wish to consider only the case when there are equal numbers of positively and negatively coupled oscillators. Aside from this constraint, we are free to choose the couplings as we wish. We consider two separate choices, which we call \textit{symmetrically} and \textit{asymmetrically} correlated disorder.


\section{Symmetrically Correlated Disorder}

\begin{figure}[!htpb]
        \includegraphics[width=0.9\linewidth]{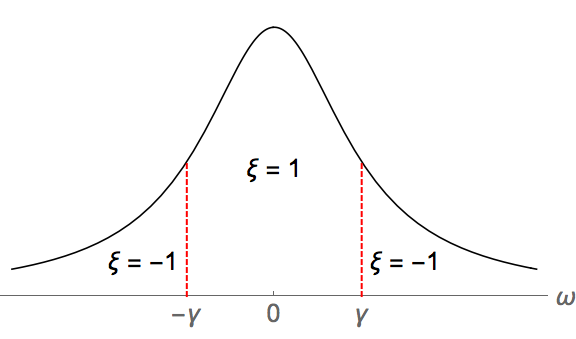}
        \caption{\label{fig:symmetric_disorder} (Color Online) Symmetrically correlated disorder. Oscillators with $\omega_i < \gamma$ and $\omega_i > \gamma$ have $\xi_i =-1$, and those with $-\gamma < \omega_i < \gamma$ have $\xi_i = +1$.}
\end{figure}

The couplings for the case of symmetrically correlated disorder are given by
\begin{equation}
\xi_i  =
\left\{
\begin{array}{ll}
-1, & i=1, \ldots, \frac{N}{4},\\
+1, & i=\frac{N}{4}+1, \ldots, \frac{3N}{4}, \\
-1, & i=\frac{3N}{4}+1, \ldots, N.
\end{array}
\right.
\label{eq:xi_sym} 
\end{equation}

\noindent
as illustrated in Fig.~\ref{fig:symmetric_disorder}. Of course, since the $\{ \omega_i \}$ and $\{ \xi_i \}$ are both deterministic functions of the index $i$, they are correlated with each other. Having chosen $\{ \omega_i \}$ and $\{ \xi_i \}$, we begin our analysis.

Collective phase synchronization is conveniently described by the complex order parameter 
\begin{equation}
Z\equiv R e^{i\Theta} = \frac{1}{N}\sum_{j=1}^N e^{i\phi_j},
\label{eq:Z}
\end{equation}

\noindent
where $R$ measures the phase coherence and $\Theta$ denotes the average phase~\cite{kuramoto84}.
Another order parameter we consider is
\begin{equation}
W\equiv S e^{i\Phi} = \frac{1}{N}\sum_{j=1}^N \xi_j e^{i\phi_j}, 
\label{eq:W}
\end{equation}

\noindent
which is a sort of weighted mean field. This quantity lets us rewrite  Eq.~\eqref{eq:model} as
\begin{equation}
\dot\phi_i = \omega_i-S\sin(\phi_i-\Phi),
\label{eq:dpdtwithW}
\end{equation}

Next we use the traditional self-consistency approach to analyze the stationary states of the Kuramoto model~\cite{kuramoto84, strogatz00}. For such stationary states, the macroscopic variables $R, \Theta, S$ and $\Phi$ are all constant in time. From Eq.~\eqref{eq:dpdtwithW} we see that the individual oscillators can exhibit two types of behavior, depending on their natural frequency. Those oscillators with $|\omega_i| \le S$ approach stable fixed points given by  $\phi_i^*=\Phi+\sin^{-1}(\omega_i/S)$, and are called ``locked." The remaining oscillators with $|\omega_i| > S$ are called ``drifters.'' They rotate nonuniformly, and have stationary density 
\begin{equation}
\rho(\phi,\omega)=\frac{\sqrt{\omega^2-S^2}}{2\pi|\omega-S\sin(\phi-\Phi)|},
\end{equation}
found from requiring $\rho \propto 1/\dot{\phi}$ and imposing the normalization condition $\int \rho \ d \phi = 1$ for all $\omega$.  

This splitting of the system into locked and drifting populations allows us to derive an expression for $W$ self-consistently:
\begin{equation}
W=\langle \xi e^{i \theta} \rangle = \langle \xi e^{i \theta} \rangle_{\rm{lock}} + \langle \xi e^{i \theta} \rangle_{\rm{drift}} .
\label{eq:self_consistency_eqn}
\end{equation}

\noindent
By going to a suitable frame, we can set $\Phi = 0 $ without loss of generality, so that $W = S e^{0} = S$. Then,
\begin{equation}
S = \langle \xi e^{i \theta} \rangle_{\rm{lock}} + \langle \xi e^{i \theta} \rangle_{\rm{drift}} .
\end{equation}

\noindent
As we vary the width $\gamma$ of the frequency distribution, the quantity $S$ will have two branches. The first of these is when only oscillators with attractive coupling $\xi_i > 0$ are locked, and the second is when oscillators with both $\xi_i > 0$ and $\xi_i < 0$ are locked.  

\subsection{First branch}

To solve for the first branch, we begin by finding the maximum frequency of the oscillators with $\xi_i >0$, which we call $\mu$. As we see in Eq.~\eqref{eq:xi_sym}, positive values of the coupling parameter $\xi_i$ are symmetrically assigned to the oscillators around $\omega=0$. Considering this, and recalling that half of the oscillators are assigned positive coupling, we determine $\mu$ from the condition that $\int_{-\mu}^{\mu} g(\omega) d\omega=\frac{1}{2}$, which then gives $\mu=\gamma$ for $g(\omega)=\frac{\gamma}{\pi}\frac{1}{\omega^2+\gamma^2}$. Thus, the oscillators with $\xi=+1$ lie in $\omega\in(-\gamma, \gamma)$, whereas the oscillators in the tails $(-\infty, -\gamma)$ and $(\gamma, \infty)$ have $\xi=-1$.

The requirement for the first branch is then $0<S\leq \mu = \gamma$. This follows from the fact that locked oscillators have $|\omega| \le S$ and that all oscillators on $(-\gamma, \gamma)$ have $\xi_i > 0$. Then the self-consistency equation~\eqref{eq:self_consistency_eqn} becomes
\begin{eqnarray}
\label{eq:S}
S  &=& \langle\xi\cos\phi\rangle_{\rm{lock}}+\langle\xi\cos\phi\rangle_{\rm{drift}} \nonumber\\
   &=& \langle (+1) \cos\phi\rangle_{|\omega|\leq S}\\
   &=& R, \nonumber 
\end{eqnarray}

\noindent
where we used $\langle\cos\phi\rangle_{\rm{drift}}=0$ due to the symmetry about $\phi=\frac{\pi}{2}$.  Equation~(\ref{eq:S}) then becomes 
\begin{equation}
S=\langle\cos\phi\rangle_{\rm{lock}} = \int_{|\omega|\leq S}d\omega~g(\omega)\sqrt{1-(\omega/S)^2}.
\label{eq:S_locked}
\end{equation}
Using $g(\omega)=\frac{\gamma}{\pi}\frac{1}{\omega^2+\gamma^2}$, the integral in 
Eq.~(\ref{eq:S_locked}) reads 
\begin{equation}
\int_{-1}^{1} dx \frac{\gamma}{\pi}\frac{\sqrt{1-x^2}}{(Sx)^2+\gamma^2}=
\frac{\pi}{\gamma}\frac{\sqrt{\gamma^2+S^2}-\gamma}{S^2}, 
\end{equation}
which gives
\begin{equation}
S=R=\sqrt{1-2\gamma}
\label{eq:R_gammac}
\end{equation}
for $\gamma\leq 1/2$. We must be careful when using  Eq.~(\ref{eq:R_gammac}), since by assumption, $S < \gamma$. In particular, Eq.~(\ref{eq:R_gammac}) is valid only for those $\gamma$ which satisfy $\sqrt{1-2\gamma} > \gamma$, which results in $\gamma > \sqrt{2} - 1 \approx 0.414$.

Equation~(\ref{eq:R_gammac}) tells us two things. The first is that there is a critical width  $\gamma_c=\frac{1}{2}$ beyond which phase coherence disappears ($S=R=0$). The second is the scaling behavior at this critical point: $S\sim (\gamma_c-\gamma)^{\beta}$ with $\beta=\frac{1}{2}$, which is same as that of conventional mean-field systems including the traditional 
Kuramoto model~\cite{kuramoto84}.


\subsection{Second branch}
On the second branch we have $S>\gamma$.  Then the oscillators with $|\omega|\leq \gamma$ have $\xi=+1$, and the other oscillators in $(-S, -\gamma)$ and $(\gamma, S)$ are locked oscillators with $\xi=-1$. The drifters still do not contribute to the phase coherence. Hence $S$ is given by 
\begin{eqnarray}
S &=&\langle \xi\cos\phi\rangle_{\rm{lock}}\nonumber\\
&=& \langle (+1) \cos\phi\rangle_{|\omega|\leq \gamma} + \langle (-1) \cos\phi\rangle_{S\geq|\omega|>\gamma} \nonumber\\
&=& 2\int_{0}^{\sin^{-1}\frac{\gamma}{S}}\cos\phi~g(S\sin\phi) S\cos\phi~d\phi \nonumber\\
&-& 2\int_{\sin^{-1}\frac{\gamma}{S}}^{\frac{\pi}{2}} \cos\phi~g(S\sin\phi) S\cos\phi~d\phi,
\end{eqnarray}

\noindent
where we used $\phi^*=\sin^{-1}\frac{\gamma}{S}$ when $\omega=\gamma$, and used $g(\omega)d\omega=g(S\sin\phi)S\cos\phi d\phi$ for the locked oscillators. The $S=0$ solution can be ignored, since we are assuming $S>\gamma$.  Thus, the second branch of partially locked states satisfies 
\begin{eqnarray}
\frac{1}{2} &=& 
\int_{0}^{\sin^{-1}\frac{\gamma}{S}}\cos^2\phi~g(S\sin\phi)~d\phi \nonumber\\
&-& \int_{\sin^{-1}\frac{\gamma}{S}}^{\pi/2}\cos^2\phi~g(S\sin\phi)~d\phi.
\label{eq:sc_S}
\end{eqnarray}

\noindent
After inserting Eq.~(\ref{eq:lorentzian}) for $g$ and evaluating the integrals using Mathematica, we find that the self-consistency equation for $S$ becomes  
\begin{eqnarray}
&&0=\pi(S^2+\gamma)-4\gamma\cos^{-1}\frac{\gamma}{S}+2\sqrt{S^2+\gamma^2} \nonumber\\
&&\times \Bigg[\sin^{-1}\Bigg(\frac{\sqrt{1-\gamma^2/S^2}}{\sqrt{2}}\Bigg)
-\sin^{-1}\Bigg(\frac{\sqrt{1+\gamma^2/S^2}}{\sqrt{2}}\Bigg)\Bigg].\nonumber\\
\label{eq:sc1}
\end{eqnarray}

\noindent
For small $\gamma$, a series solution of Eq.~(\ref{eq:sc1}) gives  
\begin{equation}
S=\sqrt{\gamma}-\frac{\gamma}{\pi}-\frac{3}{2}\frac{\gamma^{3/2}}{\pi^2}+\frac{(-24+\pi^2)\gamma^2}{6\pi^3}
+{\cal{O}}(\gamma^{5/2}).
\label{eq:S_forsmallgamma}
\end{equation}

\subsection{Comparison with simulations} 

Our results above yield analytical predictions for both branches of $S$, given by  Eq.~(\ref{eq:R_gammac}) and Eq.~(\ref{eq:sc1}), respectively. Since Eq.~(\ref{eq:sc1}) is implicit and can only be solved numerically, it is also instructive to compare the series approximation~(\ref{eq:S_forsmallgamma}) to numerical results. We performed those numerical computations in two ways: by integrating Eq.~(\ref{eq:model}) using the fourth-order Runge-Kutta (RK4) method for different $\gamma$, and by solving the self-consistency equation~(\ref{eq:sc1}) using Newton's method. 

\begin{figure}[!htpb]
        \includegraphics[width=0.95\linewidth]{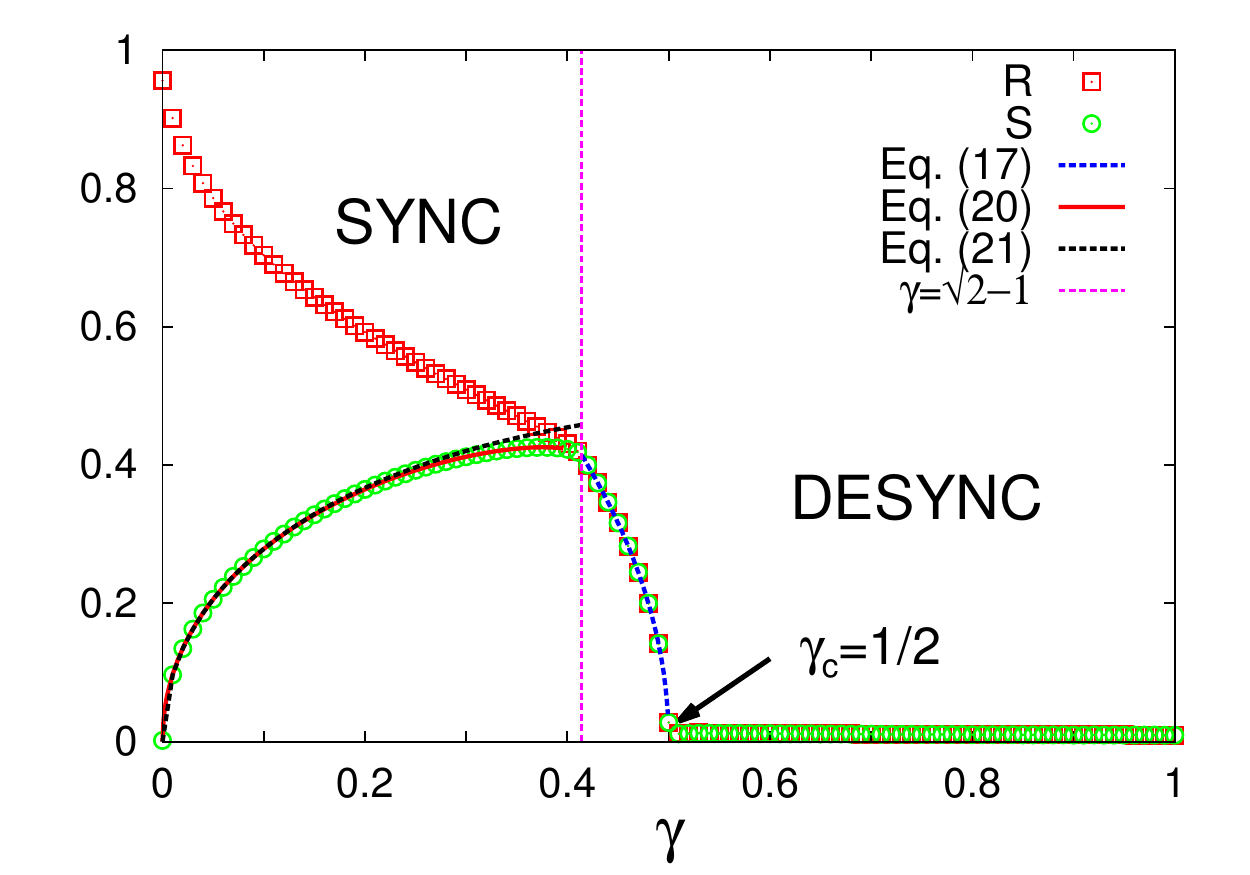}
        \caption{\label{fig:R_S_sym} 
(Color online) Behavior of the order parameters $R$ and $S$, 
represented by red open boxes and green open circles, is shown as a function of $\gamma$. 
The system size is $N=12800$, and the data have been averaged over 10 
samples with different initial conditions $\{\phi_i(0)\}$, where the 
errors (not shown) are smaller than the symbol size.  
The labels ``SYNC'' and ``DESYNC'' represent the phase coherent state and 
desynchronized state, respectively. 
The blue dashed line and red solid line represent Eq.~(\ref{eq:R_gammac}) and (\ref{eq:sc1}), 
respectively, showing a good consistency with the numerical data. 
The black dashed line represents Eq.~(\ref{eq:S_forsmallgamma}), displaying a good agreement 
with the numerical data for small $\gamma$.  The pink dashed line shows the analytical prediction $\gamma=\sqrt{2}-1$ for where the two branches join. 
}
\end{figure}

Figure~\ref{fig:R_S_sym} shows good agreement between theory and numerics for both branches of $S$. The coherent state with $R>0$ and $S>0$ emerges for $\gamma<\gamma_c=1/2$, as predicted by the self-consistency equation. (Curiously, the values of $R$ coincide exactly with those obtained from the original Kuramoto model \cite{kuramoto84} where all oscillators have the same positive coupling $\xi=+1$ for all $i$.)


The above findings answer our first question: a symmetric disorder distribution  with $\langle \omega_i \rangle=0$ and $\langle \xi_i \rangle=0$ does not necessarily prohibit phase coherence. As these results show, if $\omega_i$ and $\xi_i$ are correlated, phase coherence is possible.


\section{Asymmetrically Correlated Disorder}

\begin{figure}[!htpb]
        \includegraphics[width=0.9\linewidth]{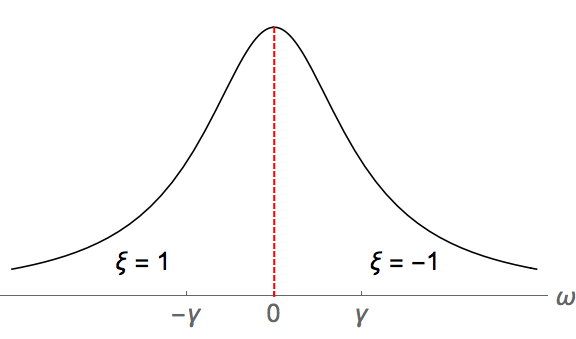}
        \caption{\label{fig:asymmetric_disorder} (Color Online) Asymmetrically correlated disorder. Oscillators with $\omega_i > 0$ have negative coupling $\xi_i < 0$, and those with $\omega_i < 0$ have positive coupling $\xi_i > 0$.}
\end{figure}

The results change considerably if the disorders are {\it{asymmetrically}} correlated.  For example, 
suppose all the oscillators with $\omega_i>0$ have a negative coupling strength $\xi_i<0$ and are therefore repulsive, while the oscillators with $\omega_i<0$ have a positive coupling strength $\xi_i>0$ and are therefore attractive: 
\begin{equation}
\xi_i  =
\left\{
\begin{array}{ll}
+1, & i=1, \ldots, \frac{N}{2},\\
-1, & i=\frac{N}{2}+1, \ldots, N, \\
\end{array}
\right.
\label{eq:xi_asym} 
\end{equation}

\noindent
as depicted in Fig.~\ref{fig:asymmetric_disorder}.


To see how the asymmetric correlation affects the phase coherence, we first study the system numerically. We simulate Eq.~(\ref{eq:model}), assuming Eqs.~(\ref{eq:regLorw}) and (\ref{eq:xi_asym}), and measure the order parameters $S$ and $R$ for various $\gamma$.


\begin{figure}[!htpb]
        \includegraphics[width=0.95\linewidth]{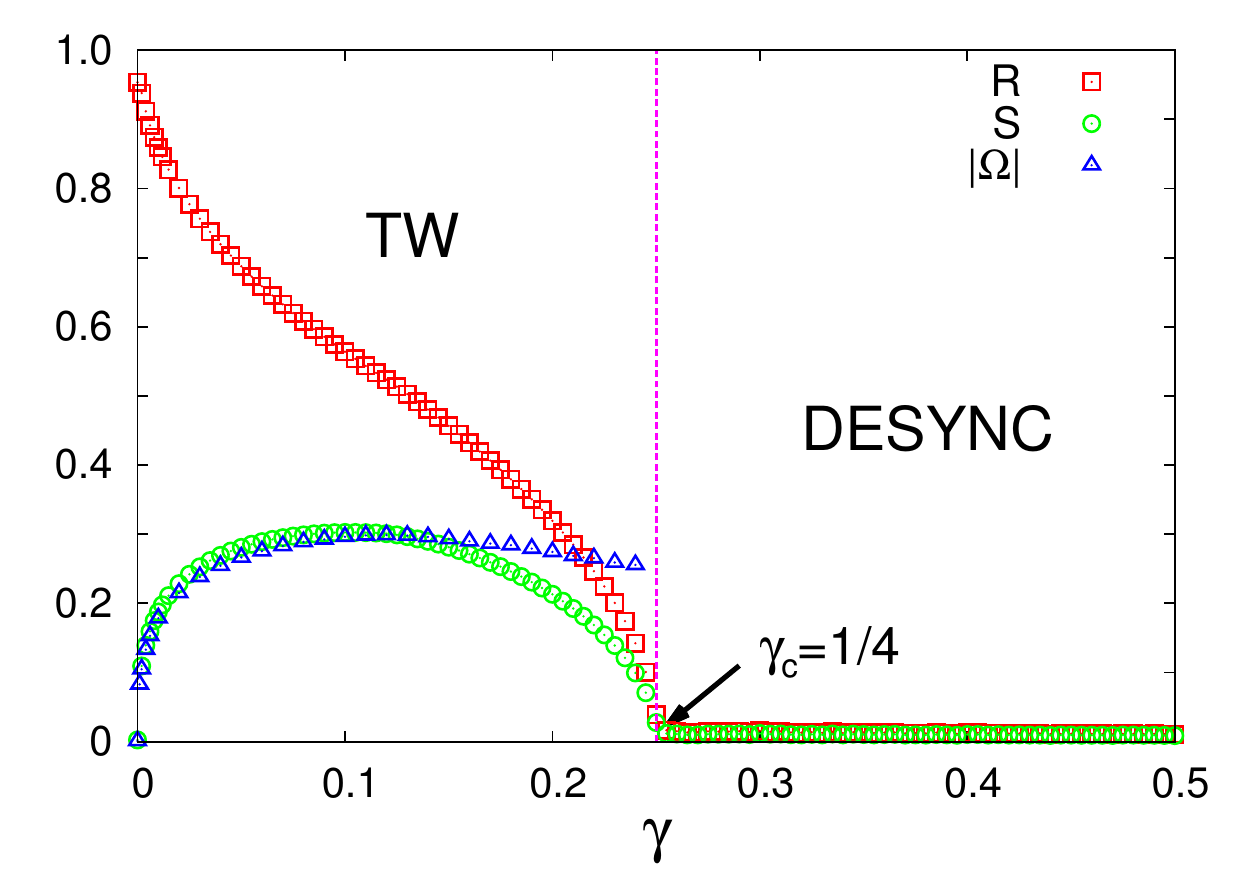}
        \caption{\label{fig:S_asym} 
(Color online) Behavior of $R$ and $S$ is shown as a function of $\gamma$ for the 
asymmetrically correlated disorders given by Eq.~(\ref{eq:regLorw}) and Eq.~(\ref{eq:xi_asym}).  The system 
size is $N=12800$, and the data are averaged over 10 samples, where the error bars (not shown) are smaller than the 
symbol size.  The label ``TW'' and ``DESYNC'' mean the traveling wave and 
desynchronized state, respectively, where the two states are separated by the black dashed line.
The absolute value of the wave speed, $|\Omega|$, shows a nonzero value for $\gamma < \gamma_c =1/4$, 
implying the emergence of the traveling wave state when $\gamma<\gamma_c$.
}
\end{figure}

We find that a phase-coherent state emerges for small $\gamma$ and  
disappears at $\gamma=\gamma_c =1/4$, as shown in Fig.~\ref{fig:S_asym}. This state, called a traveling wave \cite{ref:HS_PRL}, is qualitatively different from the coherent state seen for symmetrically correlated disorder. Instead of remaining constant, the complex order parameters $Z(t)$ and $W(t)$ each trace a circle about the origin at constant angular velocity $\Omega$ in their respective complex planes.  Or equivalently, the mean phase velocity $\langle\dot\phi(t)\rangle$ is nonzero, despite the fact that the natural frequencies satisfy  $\langle\omega\rangle=0$. 


\begin{figure}[!htpb]
        \includegraphics[width=0.48\linewidth]{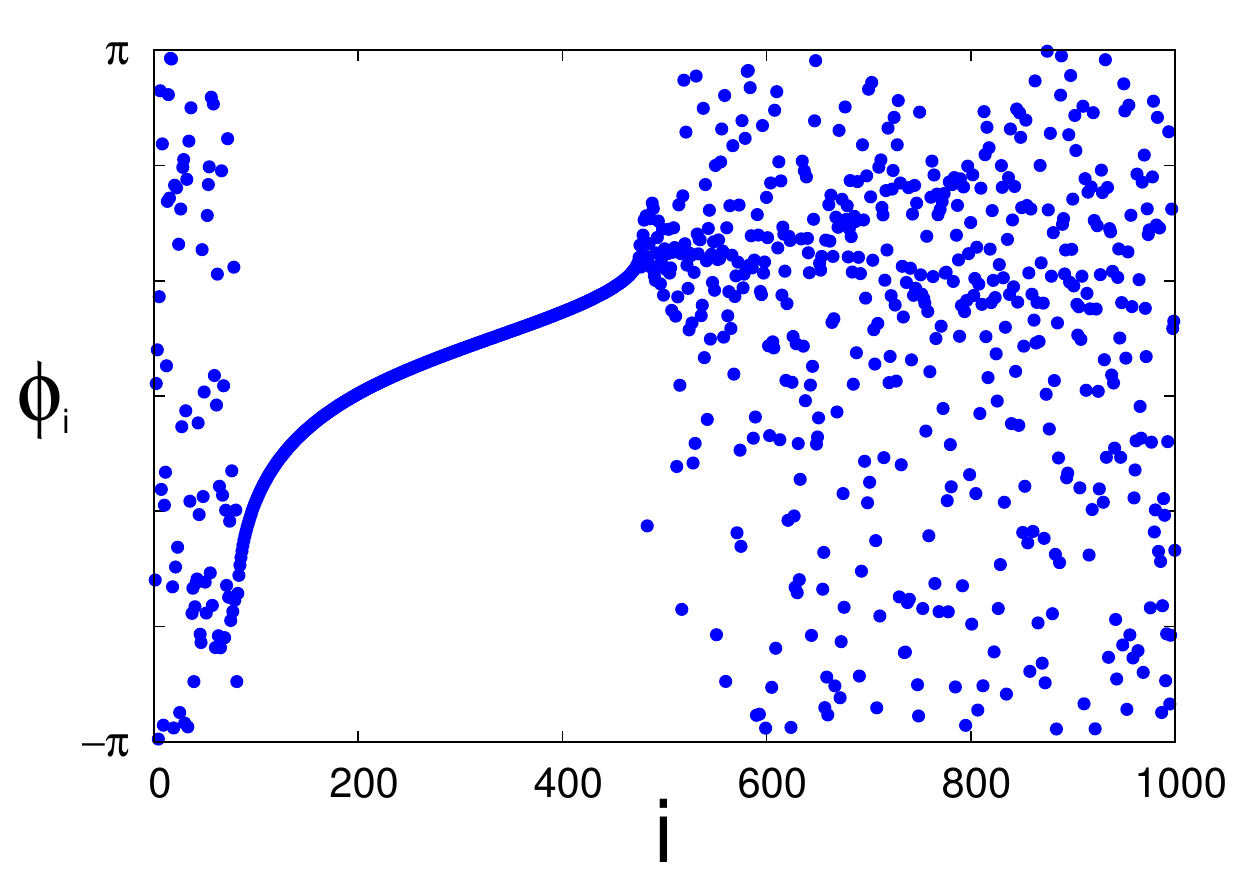}
        \includegraphics[width=0.48\linewidth]{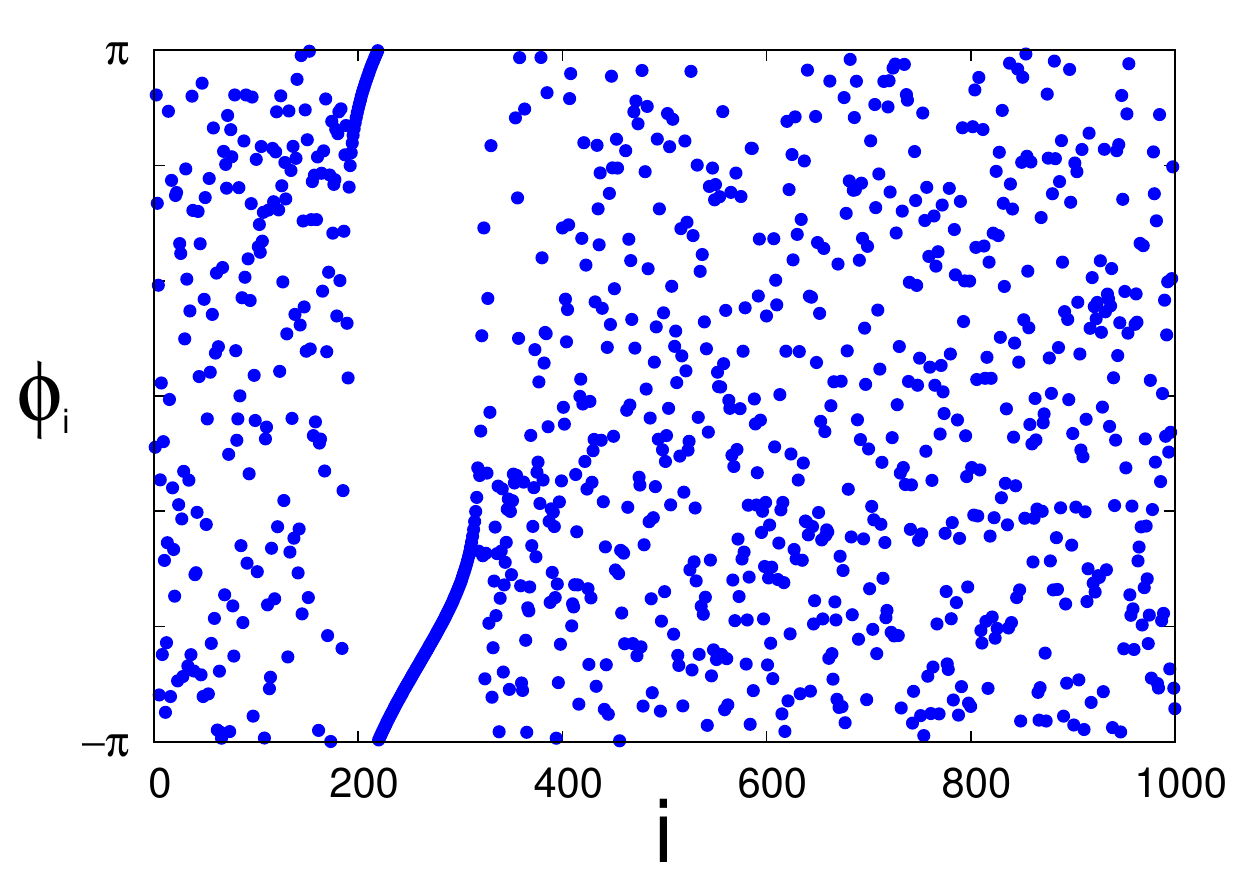}
        \includegraphics[width=0.48\linewidth]{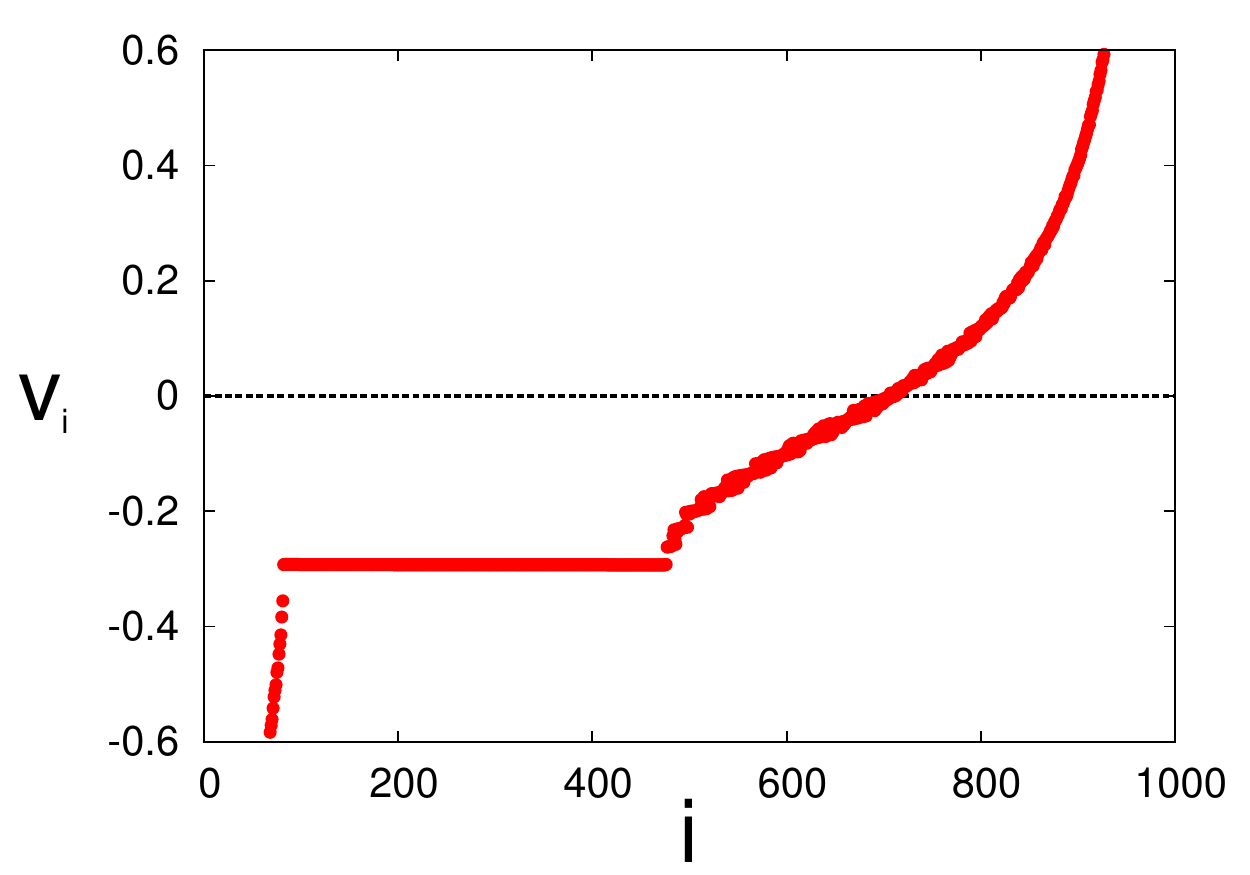}
        \includegraphics[width=0.48\linewidth]{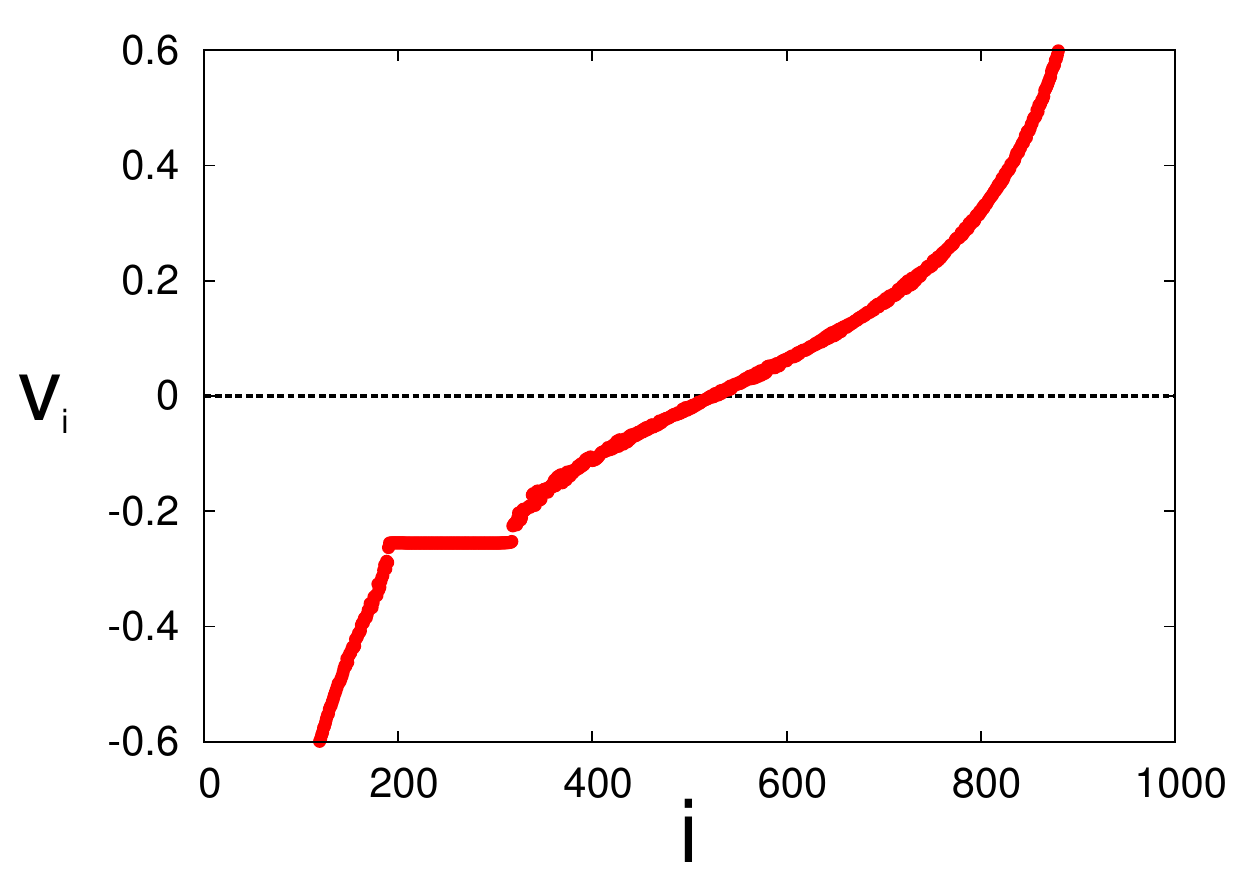}
        \caption{\label{fig:dpdt_regLorw_regdeltaK} 
(Color online) Snapshots of a traveling wave state. 
Phase $\phi_i$ (top panel) and average velocity $v_i =\langle\dot\phi_i\rangle$ (bottom panel) are 
shown for $N=1000$ oscillators for $\gamma=0.15$ (left) 
and for $\gamma=0.24$ (right), respectively. 
}
\end{figure}


To further investigate the traveling wave state, we measured the phase $\phi_i$ and the mean velocity $v_i\equiv\langle \dot\phi_i\rangle$ 
for different $\gamma$.  Figure~\ref{fig:dpdt_regLorw_regdeltaK} shows snapshots of $\phi_i$ and $v_i$ for $\gamma=0.15$ and $0.24$. Notice that in both cases,  a velocity plateau occurs for a certain range of oscillators with $\omega_i<0$ and $\xi_i > 0$; for these oscillators, the phases $\phi_i$ become locked and all the oscillators run at the same instantaneous frequency. Interestingly, the velocity plateau occurs at a nonzero value ($v_i \neq 0$), and shrinks in width as $\gamma$ is increased, until it disappears at $\gamma=1/4$.

We also measured the rotation speed of the order parameter, defined by 
\begin{equation}
\Omega = \frac{d}{dt} \arg W(t), 
\end{equation}
to further characterize the traveling wave. The absolute value of the wave speed, $|\Omega|$, is shown as a function of $\gamma$ in Fig.~\ref{fig:S_asym}. As can be seen, $|\Omega|$ is non-zero for $\gamma < 1/4$, which is consistent with the above finding of $\gamma_c = 1/4$.


\subsection{Analysis of the traveling wave state}

To derive the transition point $\gamma_c=1/4$ analytically, we again use self-consistency arguments. We first make the coordinate change $\phi \rightarrow \phi + \Omega t$, so that each oscillator obeys $\dot{\phi_i} = \omega_i - \Omega -S \sin \phi_i$. By definition of $\Omega,$ in this frame the order parameter $W$ becomes stationary and so $S$ is constant. Hence the population of oscillators again splits into locked and drifting sub-populations. The locked oscillators (in this rotating frame) are those with $\dot{\phi} = 0$, namely those with $\Omega -S \leq \omega \leq \Omega + S $. The remaining oscillators are drifters, and have either $\omega < \Omega -S $ or $\omega > \Omega + S$. The self-consistency equations then read
\begin{align}
S &= \langle \xi \cos{\theta} \rangle = \langle \xi \cos{\theta} \rangle_{\rm{lock}} \label{sc_asymmetric_disorder1}
  \\
0 &= \langle \xi \sin{ \theta} \rangle =  \langle \xi \sin{ \theta} \rangle_{\rm{lock}} +  \langle \xi \sin{ \theta}\rangle_{\rm{drift}},
\label{sc_asymmetric_disorder2}
\end{align}
where, for the same reasons as before, the drifting oscillators make zero contribution to $S$ in Eq.~\eqref{sc_asymmetric_disorder1}.

Our task is to solve these two equations simultaneously for the unknown $S$ and $\Omega$. The presence of two unknowns makes the self-consistency analysis much more difficult than it was for the case of symmetrically correlated disorder, which had just $S$ as an unknown (since the phase-coherent solution automatically had $\Omega \equiv 0$). 

As before, $S$ will have two branches (and accordingly, so will $\Omega$). On the first of these branches, only oscillators with attractive coupling $\xi > 0 $ can be locked; on the second, oscillators with $\xi < 0$ or $\xi>0$ can be locked. In terms of natural frequencies, branch 1 is defined by $ \max \omega_{\rm{drift}} < 0 \Rightarrow \Omega + S < 0,$ while branch 2 is defined by  $ \max \omega_{\rm{drift}} > 0 \Rightarrow \Omega + S > 0.$

Given the messiness of the analysis, we focus our analysis on the simpler and more interesting branch,  namely branch 1, which bifurcates from $\gamma_c=1/4$. In this case, equations \eqref{sc_asymmetric_disorder1} and\eqref{sc_asymmetric_disorder2} become 
\begin{align*}
S &= \int_{-\pi}^{\pi} \int_{\Omega - S}^{\Omega + S} \cos \phi \; g(\omega) \rho(\phi, \omega)_{\rm{lock}}  \; d \omega d \phi, \\
0 &= \int_{-\pi}^{\pi} \int_{-\infty}^{\Omega - S} \sin \phi \;  g(\omega) \rho(\phi, \omega)_{\rm{drift}}  \; d \omega d \phi \\
 &+ \int_{-\pi}^{\pi} \int_{\Omega - S}^{\Omega + S} \sin \phi \; g(\omega) \rho(\phi, \omega)_{\rm{lock}} \; d \omega d \phi \\
&+ \int_{-\pi}^{\pi} \int_{\Omega + S}^{0} \sin \phi \; g(\omega) \rho(\phi, \omega)_{\rm{drift}} \; d \omega d \phi \\
 &- \int_{-\pi}^{\pi} \int_{0}^{\infty} \sin \phi \;  g(\omega) \rho(\phi, \omega)_{\rm{drift}}  \; d \omega d \phi, \\
\end{align*}
where 
\begin{equation}
\rho(\phi, \omega)_{\rm{drift}} =\frac{\sqrt{(\omega-\Omega)^2-S^2}}{2\pi | \omega-S\sin(\phi-\Phi) | }
\end{equation}
and 
\begin{equation}
\rho( \phi, \omega)_{\rm{lock}} = \delta \left( \sin \phi- (\omega - \Omega) / S \right). 
\end{equation}

\begin{figure}[!htpb]
        \includegraphics[width=0.95\linewidth]{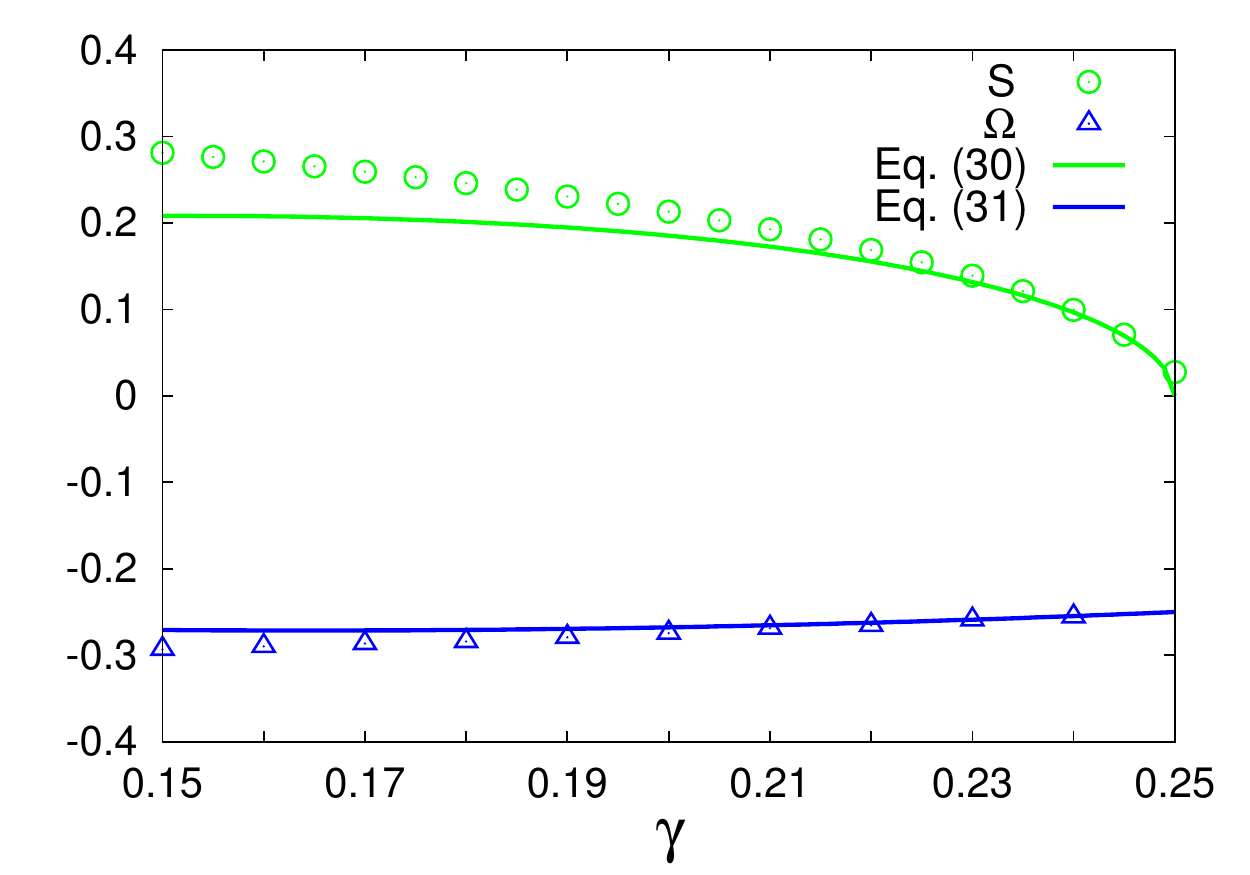}
        \caption{\label{fig:power_series_S_Omega} 
(Color online) Comparison of small-$\gamma $ expansions of $S$ and $\Omega$, as per equations \eqref{S_power_series} and \eqref{Omega_power_series}, with numerical solutions obtained from simulations of $N =12800$ oscillators. The  data are averaged over 10 samples for $S$, and one sample for $\Omega$. 
}
\end{figure}

Computing the integrals using Mathematica yields two complicated equations, $F_1(\gamma, S, \Omega) = 0$ and $F_2(\gamma, S, \Omega) = 0$, which are not enlightening to write here.  We used these equations to find power series solutions for $S$ and $\Omega$, as follows.  We first substituted the ansatz 
\begin{equation} 
S = a_1 \left( \frac{1}{4} - \gamma \right)^{1/2} + a_2 \left(\frac{1}{4} - \gamma\right) + \mathcal{O}\left(\frac{1}{4} - \gamma\right)^{3/2}
\end{equation}
and 
\begin{equation} 
\Omega = -\frac{1}{4} + b_1\left(\gamma - \frac{1}{4}\right) + b_2 \left(\frac{1}{4} - \gamma\right)^2 + \mathcal{O}\left(\frac{1}{4} - \gamma\right)^3
\end{equation}
into $F_1(\gamma, S, \Omega)=0$ and $F_2(\gamma, S, \Omega)=0$. The particular forms of these ansatzes were motivated by numerics. We then expanded the equations for small $(\frac{1}{4} - \gamma)$, and solved for constants $a_1, a_2, b_1, b_2$ by demanding that the coefficients of each power of $(\frac{1}{4} - \gamma)$ be zero. The resulting series for $S$ and $\Omega$ are given by 
\begin{align}
S &\sim \left( \frac{1}{4}-\gamma \right)^{1/2} - \left(2 + \sqrt{2}\right) \left(\frac{1}{4}-\gamma \right)^{3/2} ,
 \label{S_power_series}  \\
\Omega & \sim -\frac{1}{4} + \frac{1}{2} \left(\gamma -\frac{1}{4}\right) + \left(\frac{3}{2}+\sqrt{2}\right) \left(\frac{1}{4}-\gamma \right)^2. \label{Omega_power_series}
\end{align}
As shown in Fig.~\ref{fig:power_series_S_Omega}, these series match well with numerics.


\section{Discussion}
\noindent
In summary, we have studied a variant of the Kuramoto model with disorder in both the natural frequencies $
\omega_i$ and the coupling strengths $\xi_i$. In previous work~\cite{ref:HS_PRE}, we considered models where $\omega_i$ and $\xi_i$ were uncorrelated. In that case, if there are equal numbers of positively and negatively coupled oscillators, the phase-coherent state does not occur and the system never synchronizes. In the present work, however, we have shown that if the $\omega_i$ and $\xi_i$ are \textit{correlated}, then synchrony \textit{can} be achieved. 

We have also shown that a traveling wave can be realized for asymmetrically correlated disorder. A similar traveling wave was found in Ref.~\cite{ref:HS_PRL} for a model of conformist and contrarian oscillators. However, that model should not be confused with the one considered here. The crucial difference is that in the present model \eqref{eq:model}, the disordered coupling appears \emph{inside} the sum as $\xi_j$, whereas in the conformist/contrarian model,  it appears \emph{outside} the sum as $\xi_i$. Intuitively, the distinction is between attracting others to \emph{you}, or being attracted to \emph{them}. In more physical terms, the model studied here imposes the same mean field $W$ on each oscillator; the effect of the disordered coupling is that different oscillators \emph{contribute} to the mean field differently via their different $\xi_j$ in Eq.~\eqref{eq:model}. In contrast, in the conformist/contrarian model, oscillators with different $\xi_i$ \emph{respond} differently to the mean field, which is what makes them either conformists (who tend to align with the mean field) or contrarians (who tend to anti-align with it).  Although the uncorrelated version of this conformist/contrarian model was studied in \cite{ref:HS_PRL}, perhaps new phenomena await discovery with the inclusion of correlated disorder in such models. 

There are many other ways to put correlated disorder into Kuramoto models~\cite{ref:iatsenko13, ref:iatsenko14}. For example, one avenue would be to introduce correlations in the \emph{sizes} of the couplings as well as their signs.  In the present paper we kept the sizes constant and disordered the signs only, via the choice of the double-delta distribution \eqref{eq:Gammaxi_delta}. It would be interesting to see if size matters, in this context at least.

\begin{acknowledgments}
This research was supported by NRF Grant No. 2015R1D1A3A01016345 and Research Base Construction Fund Support Program funded by Chonbuk National University in 2015 to H.H, and NSF grant DMS1513179 to S.H.S.
\end{acknowledgments}


\end{document}